\documentclass[aps,prl,twocolumn,showpacs]{revtex4}

\usepackage{graphicx}

\def\pra{Physical Review A}

\begin{document}
\title{Optimized production of large Bose Einstein Condensates}
\author{D. Comparat\footnote{email: Daniel.Comparat@lac.u-psud.fr}, A. Fioretti, G. Stern, E. Dimova, B. Laburthe Tolra$^{\dag }$  and P. Pillet}
\affiliation{Laboratoire Aim\'{e} Cotton, CNRS II, B\^{a}t. 505, Campus  d'Orsay, 91405 Orsay cedex, France}
\affiliation{$^{\dag }$ Laboratoire de Physique des Lasers, UMR 7538 CNRS, Universit\'e de Paris 13, 99 Avenue J.-B. Cl\'ement, 93430 Villetaneuse, France}
\begin{abstract}
We suggest different simple schemes to efficiently load and evaporate a ''dimple'' crossed dipolar trap. The collisional processes between atoms which are trapped in a reservoir load in a non adiabatic way the dimple.  The reservoir trap can be provided either by a dark SPOT Magneto Optical Trap, the (aberrated)  laser beam itself or by a quadrupolar or quadratic magnetic trap.
Optimal parameters for the dimple are derived from thermodynamical equations and from loading time, including possible inelastic and Majorana losses. 
We suggest to load at relatively high temperature a tight optical trap.
Simple evaporative cooling equations, taking into account gravity, the possible occurrence of hydrodynamical regime, Feshbach resonance processes and three body recombination events are given. To have an efficient evaporation the elastic collisional rate (in s$^{-1}$) is found to be on the order of the trapping frequency and lower than one hundred  times the temperature in micro-Kelvin.    Bose Einstein condensates with more than $10^7$ atoms should be obtained in much less than one second starting from an usual MOT setup.
\end{abstract}


\date{\today}

\maketitle

The creation of degenerate atomic and molecular gases is one of the major achievements in the last decade in physics. Using a Magneto Optical Trap (MOT) laser cooled sample as the starting point, the largest Bose Einstein Condensate (BEC) ever created  contains $5\times10^7$ atoms \cite{2001Sci...292..476A,2005PhRvA..71c3617N,2005cond.mat..7348S}. The hydrogen BEC contains $10^9$ atoms but starts from a cryogenically cooled sample \cite{1998PhRvL..81.3811F}.
The techniques to form a degenerate BEC or Fermi gas have evolved but 
it is still challenging to reach quantum degeneracy in less than one second with large atom numbers. Some cooling methods such as
superposed laser cooling \cite{2000PhRvA..62a3406K,2000PhRvA..61f1403I,2001PhRvA..63a1401D},
Doppler cooling on forbidden transitions \cite{2001PhRvL..87l3002B} 
 or degenerate Raman sideband cooling using for instance a
 three-dimensional far-off-resonant lattice  have reached high phase-space density than MOTs  but with complex setups \cite{2000PhRvL..84..439K,2000PhRvL..85..724H,2001PhRvA..63e1401T,2001PhRvA..63b3405H}.

Up to now, the only way to reach the quantum degeneracy regime is to load atoms into conservative traps and perform evaporative cooling. This requires complex setups, both in terms of vacuum and in terms of optical access. The most common traps are magnetic traps, in which BEC can be achieved after tens of seconds of forced evaporation.  The use of atom chips provides a much tighter trap confinement than usual magnetic traps. Consequently evaporative cooling is much more efficient,  leading to BEC in less than a second but with atom numbers on the order of $10^5$  \cite{2001Natur.413..498H}. 
Another common trap is the optical trap.
The use of CO$_2$ dipole traps (also called QUEST trap) has led to large samples of degenerate gases of $5 \times 10^5$ atoms in few seconds starting from a MOT \cite{2001PhRvL..87a0404B,2002PhRvL..88l0405G,2003PhRvL..91x0408C,2003ApPhB..77..773C}. 
Finally combinations of traps have been partially studied. For instance
 the powerful so called 'dimple' trick, has been demonstrated \cite{1998PhRvL..81.2194S}: a small but tight trap is superimposed on a large  atomic trap acting as a temperature reservoir. For a loading efficiency of nearly $20\,$\% of the initial number of atoms, the temperature remains almost unchanged and a large gain in the atomic density and in phase space density can be obtained, allowing fast evaporative cooling. 

Several theoretical studies of evaporative cooling exist (see \cite{Ketterle1996b,2004PhRvA..70a3404T} and references therein) but none of them simultaneously takes into account gravity, the hydrodynamical regime effects, the ability to tune the scattering length value $a$ through a Feshbach resonance, the temperature dependence of the scattering cross section $\sigma $, three body recombination processes and the trap shape evolution.
In this paper, we present a study of combined traps as well as a theoretical model which takes into account all these effects. This provides tools that help devising the best strategy to reach quantum degeneracy, with many atoms, in a short time, and with a simple experimental setup. We will see for example that a simple magnetic plus crossed dipole trap setup can lead in much less than one second to a large ($>10^7$ atoms) BEC from a standard vapor cell MOT setup.

In the first part of this paper, we study loading strategies for a dipole trap from different types of reservoir traps. We mainly study the case of a dimple dipole trap focussed on a large magnetic trap. Dark-SPontaneous-force Optical Trap (dark-SPOT) \cite{1993PhRvL..70.2253K} or degraded laser beam could also be used as a reservoir to load atoms into a dimple optical trap. We also derive thermodynamical equations for the loading process. We discuss the dynamics of the loading process of the dimple, as well as the effect of Majorana losses and of two and three body inelastic losses. We emphasize that loading diabatically the dimple trap is (three times) less time consuming than trying to load it adiabatically, and that it leads more rapidly to high phase-space densities needed for efficient evaporative cooling.

The last part of this paper is devoted to the derivation of simple differential equations for evaporative cooling. We include gravity, hydrodynamical regime, the possibility of modifying $a$ with a Feshbach resonance, the temperature dependence of the scattering cross section, three body recombination and possible trap shape evolution in our model. Strategies for reaching quantum degeneracy are discussed.

 \section{Efficient loading of an optical trap}

Our goal is to provide a setup as simple as possible to rapidly reach quantum degeneracy by evaporative cooling. The speed of evaporation is directly linked to the trap frequency.  If we want to avoid to work with atom chips the solution to have a tight trap is to use an optical trap, which also allows a better optical access for further experiments on the BEC or on the degenerate Fermi gases. In this article, we mainly focus on the loading and evaporative processes in a (crossed) optical trap. As mentioned in the introduction one of the most powerful techniques to load an optical trap is to superimpose it on a large reservoir of cold atoms and wait for the collisions to fill the ''dimple'' formed by the optical trap. Once the dimple is loaded the reservoir trap is removed and the evaporation process is achieved by lowering the laser intensity to reach the quantum degeneracy regime.

Before we describe in detail the case where the atomic reservoir is  a magnetic trap, we would like to briefly discuss other possible reservoirs, such as a  dark-SPOT, and an aberrated optical trap. To use a dark-SPOT as the reservoir of atoms is an attractive idea for two reasons: first the atomic density is higher than in magneto-optical traps \cite{1993PhRvL..70.2253K,2004OptCo.235..333M}. Second, at least for alkaly atoms, atoms are in the lowest hyperfine state, in which inelastic collisions are less likely, and they barely see the MOT light. Pioneer works have been done in a non degenenerate case \cite{2003OptL...28.1266N,2005JPhB...38.1381L}, but very recently the group of Paul Lett \cite{Lett} at NIST reached quantum degeneracy with sodium atoms after loading a crossed dipole trap from a dark-SPOT. We believe that these results are due to efficient dimple loading directly from the dark-SPOT. In the experiment no depumping light is used, because the dark SPOT alone is specially efficient for sodium \cite{1994PhRvA..50.3597A,1996PhRvA..53.1702T}, but we believe this might be needed for other atoms.

A crossed optical trap alone (i.e. after that the MOT has been turned off) can also play the role of reservoir. Indeed,
the  efficiency of the loading process in a CO$_2$ laser cross  trap has been clarified by the work on Ytterbiums atoms \cite{2003PhRvL..90b3003T,2003PhRvL..91d0404T} indicating that the atoms, trapped by the  strong restoring longitudinal force existing in a CO$_2$ trap,  become concentrated into the cross 'dimple' region by atom-atom collisions. 
 The CO$_2$ laser is very efficient because it diverges more rapidly (for similar waist) than smaller wavelength laser beams (Nd:YAG for instance), forming a  longitudinal trap additionally to the radial one. The crossed region acts as a dimple in  the reservoir formed by the non crossed region.
 The main drawback of using a CO$_2$ laser is that it requires special windows (such as ZnSe one) in the setup. 
 From the waist propagation formula $w(z) = w_0 \sqrt{1 + \left( \frac{\lambda M^2 z}{\pi w_0^2} \right)^2}$, where the minimum waist is $w_0=\frac{ \lambda M^2 f}{\pi  W}$ (created by a $f$ focal length lens and a collimated laser beam of waist $W$), we see that a Far Off Resonance Trap (FORT)  laser ($\lambda \approx 1.06\,\mu$m) such as a Nd:YAG or an Yb fiber laser with
$M^2=10$  create the same trapping potential shape as a TEM$_{00}$ CO$_2$ laser   ($\lambda = 10.6\,\mu$m) with
$M^2=1$. The near and far field beam shape of a CO$_2$ $M^2=1$ laser cannot be both perfectly matched  by a near infrared $M^2=10$ laser  but this is probably a small effect and is beyond the scope of our article.
We then suggest that a CO$_2$ laser could be replaced by a $M^2\approx 10$ infrared laser, such as a
nearly gaussian shaped multimode (fiber) laser  or a diode laser. The trapping potential depth can  be matched by using appropriate laser power.
 With similar trapping potential, if   keeping a negligible photon scattering, the efficiency of the loading and of the evaporative cooling should then be the same in both laser setup but with the simplicity of using a laser light non absorbed by the glass cell. 
 
Another possible way is to use a time averaged optical trap in a similar way to the one described in reference \cite{2005PhRvA..72b3411A}. For instance by rapidly 
modulating the frequency driving an Acousto-Optic Modulator and thus by
sweeping the position of the first order diffracted beam in front of a focusing lens it should be possible to increase the trapping volume of the reservoir without changing significantly the waist size at the lens focus neither its location (which defines the dimple). 
This time-averaged trap can also be used to optimize the spatial intensity shape during the evaporative cooling process. Finally, it should be possible to use  aberrated laser beams (e.g. by computer-addressed holograms)  \cite{2005NJPh....7....4A,2005JOptA...7S.392S}.

We now turn to the use of a magnetic trap as a reservoir to load an optical trap, which is the focus of our study. 
 Magnetic traps are the largest available traps with more than $10^9$ atoms at $T\approx 100\,\mu$K  directly transferred from a MOT \cite{2005cond.mat..7348S,Lewandowski}. Superconducting magnets can even catch, but from a buffer gas cooled,
 $10^{12}$ atoms at  $600$~mK  \cite{2005PhRvA..71b5602N}.
 Recently, in reference \cite{2005cond.mat..8423G}, a Cr BEC has been obtained by using such a transfer technique from a magnetic to a crossed optical trap. Unfortunatly, in this experiment the depth of the optical trap  was very small and, even after a RF magnetic trap evaporative cooling step, it was (in temperature units) only twice the atomic temperature. This led to very fast decay due to plain evaporation after removing the magnetic trap.

\begin{figure}
\resizebox{0.45\textwidth}{!}{
\includegraphics*[0mm,0mm][127mm,145mm]{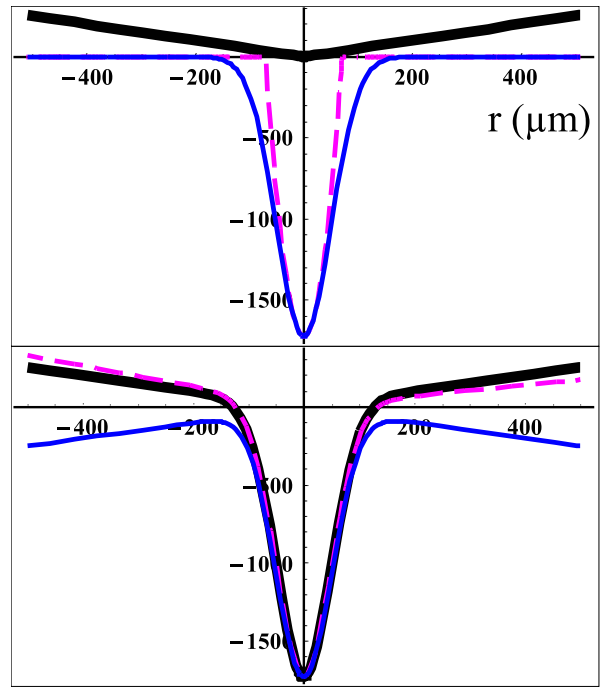}
}
\caption{Trapping potentials (in $\mu$K) for the cesium $f=3,m_f=-3$ atom. 
Top: Magnetic (thick black  solid line)  and crossed optical   (thin  blue solid line) potentials and its quadratic approximation (thin magenta dashed line). The magnetic gradient is $10\,$mT/cm. The one beam laser is a $100\,$Watt Nd:YAG laser focus on a $w_0 = 100\, \mu$m waist.
Bottom: Full potential (thick black  solid line), with  gravity included
(thin magenta dashed line). The full potential for the magnetically untrapped state $f=3,m_f=+3$ without gravity is plotted in a thin blue solid line.
}
\label{fig:pot}
\end{figure}

 The goal of this article is  to provide an optical trap with good loading and good starting condition before starting evaporative cooling. 
For simplicity reason, we will treat the case where two orthogonal laser beams cross in a central region, which is 
assumed to form a radial isotropic gaussian trap potential $
 U(r)=U_{\rm laser} =-U_0 e^{-2 r^2/w_0^2}$.  The potential depth
$U_0$ is then proportional to the one beam laser intensity $I=\frac{2P}{\pi w_0^2}$ where $P$ is the laser power. 
  With these simplifications the trapping angular frequency $\omega$ verifies $\omega =\sqrt{\frac{4 U_0}{m w_0^2}}$ where $m$ is the atom mass ($\omega =\sqrt{\frac{4 U_0 2^{1/3}}{m w_0^2}}$ in the real physical crossed dipole trap situation of identical but orthogonal laser beams).
  The experimental parameters:  $P$ and $w_0$ are related by $U_0 = \eta k_B T \propto P/w_0^2 \propto \omega^2 w_0^2$ where
 $\eta=U_0/k_B T$ is a dimensionless parameter, $k_B$ is the Boltzmann constant and $T$ the temperature  of the trapped sample.
We  illustrate the full potential (quadrupolar trap plus optical trap) in figure \ref{fig:pot} for an  optical trap with $U_0=k_B \times 1720 \mu$K and $\omega=2\pi \times 1050\,$Hz. We will illustrate most of our results using the cesium atom 
 ($f=3,m_f=-3$ which has a  magnetic moment $\mu\approx -3 \mu_B/4>0$, where $\mu_B<0$ is the Bohr magneton)
because it 
 is known to accumulate several problems (mainly due the high value of its scattering length $a=-3000\,a_0$) \cite{2003Sci...299..232W} such as hydrodynamical regime \cite{2004JPhB...37.3187M}, large 
 two body relaxation rate  \cite{1998EL.....44...25G} and large
 three body relaxation rate \cite{2003PhRvL..91l3201W}. In the cesium case the optical potential in figure \ref{fig:pot} can be created by crossing two Nd:YAG laser beams of 
  $100\,$Watt each  focused on a $w_0 = 100 \mu$m waist.

\subsection{Thermodynamics in an optical trap}

In order to study the loading of the dimple trap we will first deduce thermodynamics quantities (such as the atom number $N$ and the temperature $T$) from standard thermodynamics equations \cite{2004PhRvA..70a1403S}. Here, we will safely neglect gravity  (see figure \ref{fig:pot}). Knowing $T,N$ and a given potential $U$, we deduce the phase space density $D =  \frac{N}{Z_1} = n_0 \Lambda^3 $ where $n_0$ is the peak atomic density
and $\Lambda = \sqrt{\frac{2  \pi \hbar^2 }{m k_B T}}$ is the thermal de Broglie wavelength. The $D(N,T,U)$ function is
calculated from
 the one atom partition function 
$ Z_1 =\Lambda^{-3} \int_0^\infty 4 \pi r^2 e^{- \frac{U(r)-U(0) }{k_B T}} d r  $ (for an infinite potential depth).
  We then deduce the
Helmoltz free energy  $ F = N k_B T (\ln D - 1)$, the entropy $S= -\frac{\partial F}{\partial T}$ and the energy $ E(N,T,U) = F + T S$
of the sample.  
For a power law potential, $U(r) =U' r^{3/\delta}$ and 
 $Z_1= \Lambda^{-3} \left( \frac{k_B T}{U'}  \right)^\delta  \frac{4\pi \delta \Gamma[\delta]}{3}$ ($\Gamma$ is the Gamma function) leading,
  in the harmonic case where $U(r) =\frac{1}{2} m \omega^2 r^2  $, to $D= N \left( \frac{\hbar \omega}{k_B T} \right)^3$ 
and $n_0 = \frac{N}{(2 \pi \sigma_r^2)^{3/2}} $  where $U(\sigma_r)  = \frac{1}{2} k_B T $.

\begin{figure}
\resizebox{0.45\textwidth}{!}{
\rotatebox[origin=rB]{270}{
\includegraphics*[11mm,157mm][159mm,284mm]{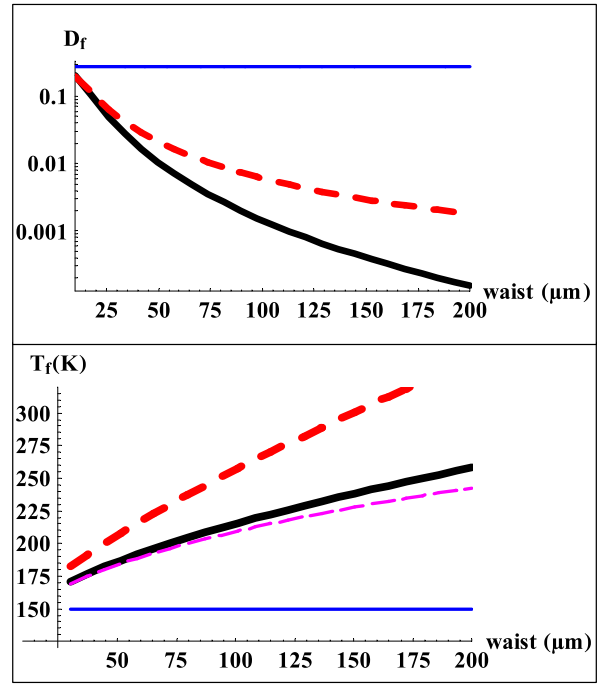}
}
}
\caption{Final phase space density (top) and temperature (bottom) at thermal equilibrium after a diabatic loading of the  optical crossed dimple gaussian trap loaded from a  reservoir containing $10^9$ atoms at temperature $150\,\mu$K. We consider different dimple waists $w_0$ but the power is adjusted to keep the dimple depth ($\eta_d\approx 11.5$) constant.  
The red thick dashed lines represent  formula (\ref{eq_ana}), the thin blue solid lines are their approximations for small $\eta_V e^{\eta_d}$ and the thick black solid lines result of our thermodynamical model, where losses are not included. 
The final temperature  after an adiabatic loading of the dimple is also indicated with a thin magenta dashed line.}
\label{fig:dimple}
\end{figure}

The atoms are assumed to be initially in 
a magnetic trap with an isotropic trapping potential $U=U_{\rm magn}$,  with initial number of atoms $N_i$, temperature $T_i$, phase space density $D_i$, entropy $S_i$ and energy $E_i= E(N_i,T_i,U_{\rm magn})$. 
 The gaussian laser trap can be  superimposed, to form
 the final potential $U_f = U_{\rm magn} +U_{\rm laser} $, either
  in an adiabatic way, i.e. with a slow change in laser power and the system evolves with constant entropy $S_f=S_i$,
 or in a sudden way with constant energy. 
 Up to now, most of the experimental and theoretical works were focused on the adiabatic process \cite{1998PhRvL..81.2194S,2004ApPhB..79.1013K,2004PhRvA..70a1403S}. However, it is faster to suddenly apply a small dimple trap and the difference in the final temperature (see figure \ref{fig:dimple}) or atom number is marginal compared to the adiabatic loading process.  In this paper, we choose to discuss the diabatic loading process.

The initial atomic spatial density distribution in the magnetic trap is $n_i (r) = (n_0)_i e^{- \frac{U_{\rm magn} (r) }{k_B T}}$, and the 
initial energy $E_i $  becomes $E_i' = E_i -  \int_0^\infty  4 \pi r^2 n_i (r) U_{\rm laser}(r) d r$ when the laser trap is suddenly added to  the magnetic trap. 
After thermalization the final energy $E_f$ of the atoms in the potential $U_f$ should be the same as the initial one $E_i'$. The magnetic and laser potentials have not the same  potential energy  at the center ($U_{\rm magn}(0)=U_{\rm laser} (\infty) =0$) then we have $E_f =E(N,T_f,U_f) + N U_f(0)$. This leads to equation $E (N,T_f,U_f) + N U_f(0) = E_i' (N,T_i,U_f)$ which is used to determine the final parameters of the sample after a sudden transformation 
(the equation in the adiabatic case is $S(N,T_f,U_f)=S(N,T_i,U_i)$) such as the final temperature $T_f$ from which we deduce the final peak density $(n_0)_f$. 
The number of transfered atoms $N_f$ left in the dimple trap after removing the magnetic trap is  $ 
N_f = \int_0^{r_d} 4 \pi r^2 (n_0)_f e^{- \frac{U_f(r)-U_f(0) }{k_B T_f}} d r
$ where the dimple radius $r_d$  is defined by by $U_f(r_d)=0$. 

Our numerical results are given in figure \ref{fig:dimple} (see also the top part of figure \ref{fig:dimple2}) where they are compared with  
 the one dimensional quadratic + two dimensional box model  developed in reference \cite{2004PhRvA..70a1403S} and that, after small calculations,  lead in our case to the following formulas:
\begin{eqnarray}  
T_f &= & T_i \left[ 1 +  \frac{\eta_d \eta_V }{2} \left( 1 +  \frac{N_f }{N_i}  \right) \right] \approx T_i \label{eq_ana}\\
\frac{N_f}{N_i} &=& \frac{\eta_V e^{-\Delta/T_f}}{1- \eta_V +\eta_V e^{-\Delta/T_f}   } \approx \eta_V e^{ \eta_d } \nonumber\\
\frac{D_f}{ D_i} & = & \frac{1}{\eta_V} \frac{N_f }{N_i} \left(  \frac{T_f }{T_i}  \right)^2 \approx e^{\eta_d } \nonumber
\end{eqnarray} 
 with $\Delta =\frac{U_f(0)}{k_B}=- \eta_d  T_i<0$      and $ \eta_V = \frac{V_d}{V_e} $ where
 $V_e=N_i/(n_0)_i={Z_1}_i \Lambda_i^3$ is the effective volume of the magnetic trap 
 ($l=V_e^{1/3}$ is the magnetic trap size)
 and
 $V_d=N_f/(n_0)_f$ is the effective dimple volume. We found that $V_d$ is typically two times $ \left(\frac{\pi w^2}{2 \eta_d}\right)^{3/2}$, where $ \left(\frac{\pi w^2}{2 \eta_d}\right)^{3/2}$ is the value calculated using the quadratic approximation of the gaussian trap. The right side approximations in equations (\ref{eq_ana}) hold when the key parameter  $ \eta_V e^{-\Delta/T_f} \approx \eta_V e^{ \eta_d }$ is small compared to 1.
As indicated by the figures \ref{fig:dimple} and \ref{fig:dimple2}, the
analytical equations (\ref{eq_ana}) inspired by the model described in reference  \cite{2004PhRvA..70a1403S} are found to be qualitatively correct and might be used as a first guide toward the optimized strategy. 

Equations (\ref{eq_ana}) seem to indicate
 that a very small $\eta_V \lesssim e^{-\eta_d}$ but deep ($\eta_d \gg 1$) dimple trap would lead to large atom number loaded in the dimple, and to a very high phase space density. 
On the contrary, we will show that it is almost impossible to reach degeneracy using only such a very deep and small dimple trap because of hydrodynamical regime, two or three body inelastic or Majorana losses.

\subsection{Dynamics  of the optical trap loading}

In this section, we give a physical picture of the mechanisms for the diabatic loading of the dimple trap. The diabatic loading relies on thermalization, and our description clarifies how collisions rapidly fill the trap. Our estimate for the diabatic loading time $t_{\rm load} $ is on the order of a few times the collision time in the magnetic trap, comparable to the estimate one given in \cite{2001PhRvA..63c3603V}. This time is smaller than the timescale for an adiabatic loading, for which the trap should be applied slowly, so that thermal equilibrium is maintained throughout the loading. In this paper we choose to study the diabatic loading which is faster than the adiabatic one, 
while  allowing similar final temperatures (see figure \ref{fig:dimple}) and still large gains in phase space densities.

\begin{figure}
\resizebox{0.45\textwidth}{!}{
\rotatebox[origin=rB]{270}{
\includegraphics*[26mm,34mm][129mm,191mm]{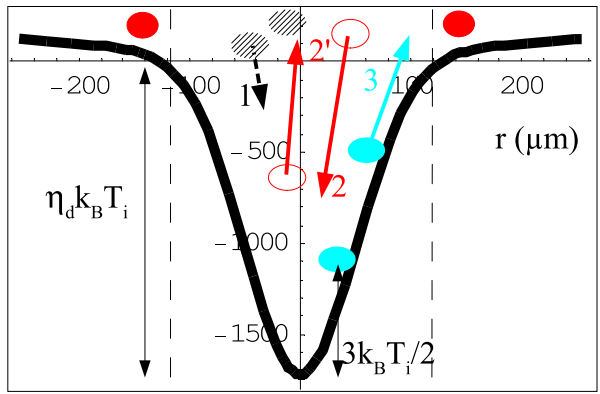}
}
}
\caption{Three types of collision occurring between atoms to load a dimple from a reservoir. The vertical dashed lines limit the $N_d$ atoms in the dimple region (optically trapped or not) from the others.
First: collision between two magnetically but not optically trapped atoms (black hatched ones) with kinetic energy  $  \approx (\eta_d + \frac{3}{2}) k_B T_i$ near the center. There is a probability $p_t  \approx 0.5$ to transfer one atom in the optical trap (process 1).
Second: between one un(-optically)trapped atom and one trapped atom (empty red  ones). There is a probability $p_{\rm ev} \approx 0.15$ to eject (evaporate) the trapped atom (process 2').
Third: between two trapped atoms (filled blue  ones) having kinetic energy  on the order of $ E_K^d = \frac{3}{2} k_B T_i$, this  thermalisation process could rarely lead to evaporation (process 3).}
\label{fig:dimple_loading}
\end{figure}

The detailed physics of the loading process is  complex and depend on three parameters that we will study one after the other:  
the probability to be transfered into the dimple (or to be  ejected)  during  a collision,  the atomic flux entering the dimple region (depending on the
collisions and oscillations times in the reservoir trap) and the collisional rate in the dimple region.
To simplify, we consider here that $\eta_d >4$ which is favorable for the subsequent evaporative cooling step and we suppose in this section
 that
the dimple trap is small enough such that  the reservoir is unaffected  during the loading: $N_f \ll N_i$ and 
$k_B T_f \approx  k_B T_i$ for instance (see figure \ref{fig:dimple}).

The first parameter (probability to be transfered or ejected) depends on the energy change occurring during a collision.
 For a Boltzmann distribution,  the probability that during a collision a given atom (with a typical kinetic energy $E_K \approx \frac{3}{2} k_B T$) acquires an energy greater than $\eta k_B T$ can be estimated to be  $p\left( \frac{\eta k_B T}{E_K} \right)= f(\eta)/f(0)=1+2e^{-\eta} \sqrt{\eta}/\sqrt{\pi} - {\rm Erf}(\sqrt{\eta})$ where $f(\eta)= \int_{\eta k_B T}^{+ \infty} \sqrt{E'_K} e^{-\frac{E'_K}{k_B T}} d E'_K $ and ${\rm Erf}$ is the error function \cite{Ketterle1996b,2004PhRvA..70e3409D}. 
Following notations of figure  \ref{fig:dimple_loading} let us first study collisions between two magnetically but not optically trapped atoms: these atoms of the reservoir have a total energy larger than the depth of the dimple trap because they have a typical energy $\left( \eta_d + \frac{3}{2} \right) k_B T_i$ while in the vicinity of $r=0$. During one of these collisions one atom is transfered into the dimple (process 1 in figure  \ref{fig:dimple_loading}) if its energy becomes less than the trapping energy $\eta_d k_B T_i$, implying  the other atom acquires an energy higher than $ (\eta_d + 3) k_B T_i$ which occurs with the probability
$ p_t = p\left( \frac{(\eta_d + 3) k_B T_i}{(\eta_d + \frac{3}{2}) k_B T_i} \right) \approx p\left( 1 \right) \approx 0.5 $. 
If one atom from the reservoir collides with an optically trapped atom,  the chance that one atom ends up trapped in the dimple is even higher (process 2 in figure \ref{fig:dimple_loading}). In order to be conservative we will assume
 $p_t \approx 0.5$ during the full loading process.

The reverse processes (2' and 3 in figure \ref{fig:dimple_loading}), namely the ejection  of an optically trapped atom after a collision, must also be taken into account.
 This probability $p_{\rm ev}$ can be evaluated in the worst case (2') when the collision occurs with an  atom from the reservoir, whose kinetic energy is $\sim \eta_d k_B T_i$. 
 The dimple atom kinetic energy is $ \sim  k_B T_i$, except during the negligible time of the very beginning of the loading   when the dimple atoms are not thermalized. This collision is then similar (in the center of mass point of view) to a collision of two atoms with $ \eta_d k_B T_i/2$ average kinetic energy inside a trap of $\eta_d k_B T_i$ depth. The collision has therefore a probability estimated to be $ p\left( \frac{\eta_d k_B T_i}{\eta_d k_B T_i/2} \right) = p(2) \approx 0.3$ 
 to leave one atom (the originally trapped or not) out of the dimple. 
 Thus, the ejection probability is $p_{\rm ev} \approx 0.15 $.  As mentioned this value is evaluated in the worst case, we then assume in the following $p_{\rm ev}\ll 1$ and neglect 
 the marginal ejection process
 during  the whole loading time.

\begin{figure}
\resizebox{0.45\textwidth}{!}{
\includegraphics*[54mm,45mm][180mm,170mm]{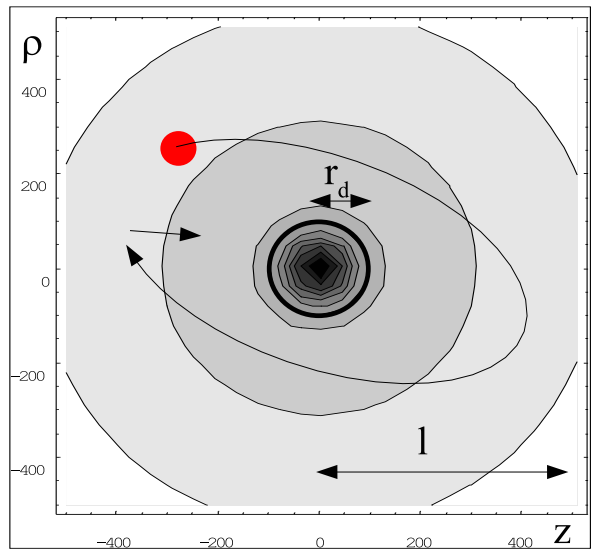}
}
\caption{Equipotential lines for the magnetic and optical potential described in figure \ref{fig:pot}. The coordinates $z$ and $\rho=\sqrt{x^2+y^2}$ are expressed in $\mu$m
(a part with the unphysical $\rho<0$ region is also drawn by symmetry for clarity). The dimple radius $r_d$ and the magnetic trap size $l$ are indicated. A typical atomic trajectory is shown with a collision occurring slightly before the oscillatory time $t_{\rm osc}$.}
\label{fig:osci}
\end{figure}

   Once the collisional processes are known we should consider the atomic flux entering in the dimple region. We use a single  notation $N_d$ for all the atoms in the dimple region (optically trapped in the dimple or not, see figure \ref{fig:dimple_loading}). At the very beginning of the loading process the $N_d \approx N_i \frac{r_d^3}{l^3} \approx n_i r_d^3$ value comes from  atoms of the reservoir and, when the sample is thermalized the $N_d \approx N_f \gg N_i \frac{r_d^3}{l^3}$  value mainly comes from 
 optically trapped  atoms.
 
Two typical times have to be taken into account to understand the dynamics of the loading. First, the ''oscillatory'' time $t_{\rm osc}$, defined by $m v_i^2 \sim  k_B T_i$ where $v_i \sim l/t_{\rm osc}$, after which an atom is almost back at the same position. This corresponds to the period in a harmonic trap. The second time is the average time interval between two collisions in the reservoir,  $t_{\rm coll} \approx \frac{1}{n_i \sigma v_i}
$ where $\sigma$ is the scattering cross section and where the exact numerical factor depends on the potential trap geometry and will be discussed in the section devoted to the evaporative cooling study.
In absence of collisions, 
a fraction  
 $\sim r_d^2/l^2$ of the magnetically trapped atoms trajectories cross the small dimple radius as shown by the figure \ref{fig:osci}
 (remember that $\rho=\sqrt{x^2+y^2}$ is in fact a two dimensional coordinate). To calculate the rate of atoms going through the dimple volume, one needs to take into account both this fraction, and the "ergodicity time" it is required to modify the trajectories. Indeed, once all the atoms whose trajectories go through the dimple are loaded into it, collisions in the reservoir are needed to modify the trajectories so that some more atoms have a trajectory going through the dimple. 
    This time is of the order  (exactly for a perfect isotropic harmonic trap)  of the collisional time  $ t_{\rm coll}$ in the magnetic trap \cite{1996PhRvA..53.3403S}. For simplicity we assume that the magnetic trap is not in a hydrodynamical regime: $t_{\rm coll}> t_{\rm osc}$, 
   which is slightly wrong in the potential we consider here (see figure \ref{fig:pot}) where
 $t_{\rm osc} = 9 \,$ms and $t_{\rm coll} = 7\,$ms. With such assumption, the number of atoms passing through the dimple region during a time $\Delta t>t_{\rm coll}$ is $\Delta N_d^{\rm pass}\sim  N_i \frac{r_d^2}{l^2} \frac{\Delta t}{t_{\rm coll}} $. 

\begin{figure}
\resizebox{0.45\textwidth}{!}{
\rotatebox[origin=rB]{270}{
\includegraphics*[6mm,47mm][196mm,176mm]{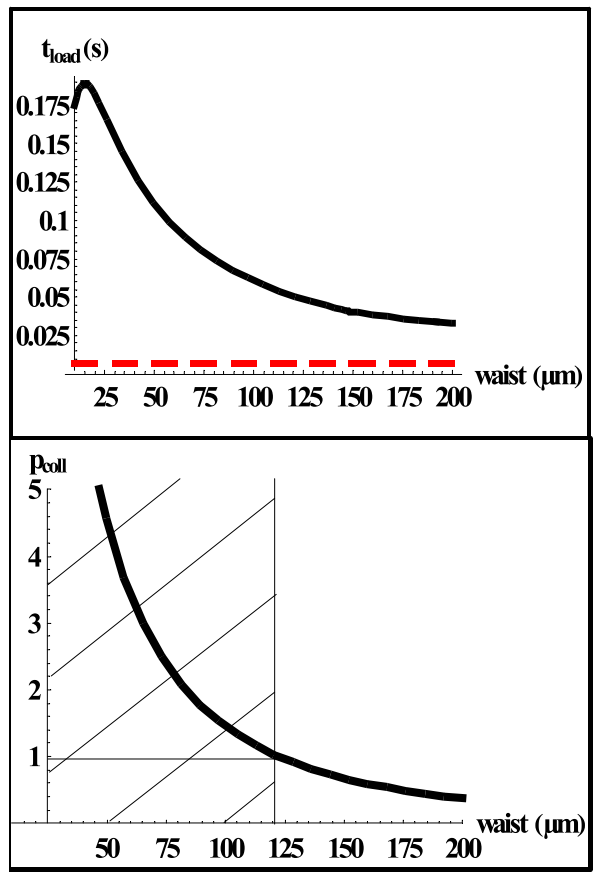}
}
} 
\caption{Dimple loading time $t_{\rm load}   $ (solid line), compared to the initial collisional time (dashed line) $t_{\rm coll}$ (here $ 7\,$ms), (top) and number of collisions before the evaporation point   $p_{\rm coll}$
 (bottom) which indicates when higher than unity the hydrodynamical regime (hatched area). The conditions in the loaded dimple trap are the same as in figure \ref{fig:dimple}:   Nd:YAG optical crossed gaussian trap (with  constant depth) loaded from a $10\,$mT/cm gradient magnetic quadrupole trap containing $10^9$ atoms at temperature $150\,\mu$K.}
\label{fig:dimple3}
\end{figure}

Each atom passing in the dimple volume has a probability $ \sim r_d (n_d \sigma)$ to collide during a single travel inside the dimple region
 and  $n_d \sim \frac{N_d}{r_d^3}$ is the density.  Before colliding with another atom from the reservoir, this atom passes through the dimple  a maximum of 
$ \frac{t_{\rm coll}}{t_{\rm osc}}$ times. Therefore,
the probability $p_d$ to transfer such an atom is then $p_d \sim  p_t r_d (n_d \sigma) \frac{t_{\rm coll}}{t_{\rm osc}}$.  Thus, the number of atoms transfered inside the dimple during $\Delta t$ might be estimated to be $\Delta N_d \sim \Delta N_d^{\rm pass} p_d$
leading to the differential equation
$ \dot N_d \sim N_d \frac{\sigma N_i}{2 t_{\rm osc} l^2 } 
$
where the dot designs the time derivative.
The loading process changes when the number of atoms inside the dimple region reaches  $N_d^0=n_d^0 r_d^3$ defined by $ r_d  \frac{N_d^0}{r_d^3} \sigma \frac{t_{\rm coll}}{t_{\rm osc}}\sim 1/p_t=2$. 
Using $t_{\rm coll} \approx \frac{1}{n_i \sigma v_i} \approx \frac{t_{\rm osc}}{n_i \sigma l}$ we found the relation
$n_d^0  = 2 \frac{l}{r_d} n_i$. 
When $N_d>N_d^0$,
 the hydrodynamical regime occurs and  any atom going through the dimple will collide with at least one atom of the dimple, the probability $p_d$ of transferring an atom is then about one each time an atom enters the dimple region. 
 The hydrodynamical regime is not favorable for evaporative cooling so we have advantage to choose a dimple such as 
$N_f \lesssim N_d^0$. In this case  the loading time of the dimple is 
\begin{equation}
t_{\rm load} \sim 2 t_{\rm osc} \frac{l^2}{\sigma N_i} \ln \left( \frac{N_f}{N_i r_d^3/l^3} \right) \sim 2 t_{\rm coll} \ln \left( \frac{n_f}{n_i} \right) \label{eq:load}
\end{equation}
 This exponential time loading is eventually followed by a linear time loading  (with $p_d =1$) if $N_f>N_d^0$.  The results for our dynamical loading model are illustrated in figure \ref{fig:dimple3}\footnote{The calculation is done as following: we exactly calculate $N_f,n_f,n_i,r_d,l$ from the exact thermodynamical model (not the analytical expressions (\ref{eq_ana})).
 We then
 calculate $t_{\rm load}$  (exponential  plus linear)  loading time: 
 $ t_{\rm load} = 2 t_{\rm coll} \ln \left( \frac{N_f/r_d^3}{N_i /l^3} \right)$ for $N_f<N_d^0$ and $
t_{\rm load} = 2 t_{\rm coll} \ln \left( \frac{N_d^0/r_d^3}{N_i/l^3} \right) + t_{\rm coll} \frac{l^2}{r_d^2} \frac{N_f - N_d^0}{N_i} $ for $N_f>N_d^0$. Where $N_d^0 =
2 \frac{t_{\rm osc}}{t_{\rm coll}}  \frac{r_d^2}{\sigma}$ with
$t_{\rm osc} = \sqrt{\frac{m l^2}{k_B T_i}}$,
$ t_{\rm coll} = 1/\Gamma_{\rm el}$ and $\sigma$  calculated using the initial magnetic trap parameters.
Finally $p_{\rm coll}=\frac{\Gamma_{\rm el}}{4 \omega}$ is calculated using the final magnetic trap parameters.}.
 We could test the accuracy of our naive theory in two ways. First using
the case of a dimple as large as the reservoir $r_d \approx l$, which is not cover by the assumptions made to derive all the previous formula, but which is nevertheless useful to test if all the initial atoms are in fact found to be transfered into the dimple (remember that $\eta_d>4$). In this case  $n_f  \approx 2 \frac{l}{r_d} n_i  $ formula reads $N_f \approx n_f r_d^3 \approx 2 n_i l^3 \approx 2 N_i$ which is only wrong by a factor 2. The second test is the fact that the hydrodynamical regime is found to be reached for the same waist $w_0 \approx 120-150\,\mu$m using our theory (see figure \ref{fig:dimple2} where $N_d^0\approx N_f$) or the exact parameter $p_{\rm coll}$  (see figure \ref{fig:dimple3}).
 It might also be useful to test here if the diabatic loading time $t_{\rm load}$  is smaller that the adiabatic one  $t_{\rm load, ad}$. If we  suppose that the adiabaticity criterion (for an harmonic trap) $\frac{\dot \omega}{\omega} \ll \Gamma_{\rm el},\frac{\omega}{2 \pi}$ is correct for a ratio value of $1/10\approx1/12$ (to simplify the final formula); this  leads, in a non hydrodynamical regime where $\Gamma_{\rm el}=\frac{1}{t_{\rm coll}}<\omega$,  to $\dot \omega = \frac{\omega }{12 t_{\rm coll}} $. Thus, $t_{\rm load, ad} \approx 12 t_{\rm coll} \ln \left( \frac{\omega_f}{\omega_i} \right)$. But $n\propto N \omega^3/T^{3/2}$ leads to $t_{\rm load, ad} \approx 4 t_{\rm coll} \ln \left( \frac{n_f}{n_i} \frac{N_i}{N_f} \right) \approx  2 t_{\rm load} \left(1 + \frac{ \ln \left( \frac{N_i}{N_f} \right) }{ \ln \left( \frac{n_f}{n_i} \right) } \right)$. A reasonable estimated of the adiabatic loading time is then $$
 t_{\rm load, ad} \sim 3 t_{\rm load}. 
 $$

   In conclusion, we found from figure \ref{fig:dimple3} that 
 a few reservoir collisional times are enough to load the dimple in reasonable agreement with the estimate of \cite{2001PhRvA..63c3603V} but we found  a strong dependence on the dimple radius $r_d$. 
 We have advantage to choose a dimple radius $r_d$ such as  the transfer atom number found by the thermodynamics equations 
$N_f$ verifies $N_f \approx N_d^0 \approx 2 \frac{r_d^2}{\sigma} $ which is the maximum value (for an almost hydrodynamical regime in the reservoir) before reaching the hydrodynamical regime in the dimple. The final density in the dimple is then $ n_f  \approx 2 \frac{l}{r_d} n_i$
reached after a loading time  $2 t_{\rm coll} \ln \left(2 \frac{l}{r_d} \right)$. For an optical dimple trap,
the link between $r_d$ and the laser waist depends on $r_d$ and on the reservoir but $r_d^2 \approx 2 w_0^2$ seems to be a reasonable approximation leading to $N_d^0 \approx 4 \frac{w_0^2}{\sigma} $ which is found (see figure \ref{fig:dimple2}) to be a very precise approximation.

 \subsection{Losses during the optical trap loading}

 We now describe how the previous results are modified in the presence of losses (inelastic losses and Majorana losses). 

We suggest to use a quadrupole magnetic trap as the reservoir
because  the trap volume is large and the oscillation frequency is high.
Then $U_{\rm magn} \approx  \mu B' r$ where $\mu$ is the atomic magnetic moment and $B'$ is the magnetic field gradient. Furthermore, with near zero magnetic field $B$ at the center, where the optical trap takes place, a quadrupole trap is a good choice to limit the two body inelastic processes. Indeed, assuming that no resonance processes, such as shape or Feshbach resonance take place,  the magnetic field dependence of the two body inelastic rate is, for the worst case, proportional to $\sqrt{B}$
 \cite{1996PhRvA..53...19M,1996PhRvA..53.1447F,1998EL.....44...25G,2002PhRvA..65e2712V,2003ApPhB..77..765H}.
 For instance in the cesium case  this rate is $ K_2 = \frac{\dot N}{N\bar n} \approx 4\times 10^{-11} B^2 (mT) T(\mu {\rm K})^{-0.78} {\rm cm}^3 {\rm s}^{-1}$ 
  where $\bar n $ is the average atomic density
 \cite{1998EL.....44...25G}. During the loading, the number $N_2$ of atoms lost due to two body collisions follows the equation
  $\dot N_2 \sim K_2 N_d   n_d \sim K_2 n_d^2 r_d^3$. Solving this equation leads to the final number $N_{2,f}$ of lost atoms due to two body collisions
  $ N_{2,f} \sim t_{\rm coll} K_2 n_i^2 e^{ \frac{t_{\rm load}}{ t_{\rm coll} }} r_d^3 \sim t_{\rm coll} K_2 n_f^2  r_d^3$. The lost fraction $\frac{N_{2,f}}{N_f}$ is then simply  $   \sim t_{\rm coll} K_2 n_f  $, i.e. the final two body loss during one collisional time (which is the only relevant time during the dimple exponential loading). Similarly the  three body atom losses (rate $\Gamma_3\sim L_3 n_d^2$) lead to a lost fraction 
  $   \sim \frac{2}{3} t_{\rm coll} L_3 n_f^2  $, i.e. roughly the final three body loss during one collisional time.

One issue with a quadrupole trap is Majorana losses, due to spin flips occurring when an atom crosses the Majorana sphere of radius $r_{\rm Maj} \sim \sqrt{\frac{\hbar v_d}{\mu B'}}$ (typically on the order of one micron) where $v_d$ is the velocity inside the dimple trap  \cite{1995PhRvL..74.3352P}.
Since the spin flip occurs near a zero magnetic field the change in potential energy is marginal and can be neglected, as shown by comparing the thin solid blue line and the thick black solid line in the bottom inset of figure \ref{fig:pot}.
  The density in the dimple trap is much higher than the one in the magnetic trap so spin flips will mainly occur from atoms already inside the dimple trap where $v_d\sim \sqrt{k_B T_f/m}$.   
  The spin flipped atoms are trapped in the optical trap. However, for these atoms, the combined magnetic plus dimple trap depth is finite (see figure \ref{fig:pot}) , which leads to evaporation at a rate $\Gamma_{\rm ev}$. This process is not necessarily  a bad one because it might be seen as the beginning of an evaporative cooling inside the dimple.  Collisions could also occur
with non optically trapped atoms with a maximum rate of $p_{\rm ev}/t_{\rm osc}$.
  Of course, another possible strategy, to combine small two body losses with negligible  Majorana losses, is to focus the crossed optical trap slightly off 
  center in a region where the magnetic field is small but not zero.
  
   To give an  estimate of the number of spin flipped atoms, we consider that no spin flip occurs through inelastic collisions. 
   This assumption holds here because for our quadrupolar trap case $K_2$ has a negligible effect (see figure \ref{fig:dimple2}).
   Following the previous reasoning concerning $\Delta N_d^{\rm pass}$ we
find that the spin flipped atom number during $\Delta t$ is roughly $N_d^{\rm Maj} \sim N_d \frac{r_{\rm Maj}^2}{r_d^2} \frac{\Delta t}{  t_{\rm coll,d}} $ where $t_{\rm coll,d} \sim t_{\rm coll} n_i/n_d $ is the collision time inside the dimple trap. This result
modifies the Majorana rate calculation given in reference \cite{1995PhRvL..74.3352P} by a factor $\frac{t_{\rm osc}}{t_{\rm coll}}$.
As a consequence,
 $\dot N_d^{\rm Maj} \sim N_d \frac{r_{\rm Maj}^2}{n_i r_d^2 t_{\rm coll} } n_d $, which is exactly the equation corresponding to two-body losses, with $K_2 $ replaced by $ \frac{r_{\rm Maj}^2}{n_i r_d^2 t_{\rm coll} }$. Thus, the final  spin flipped fraction  is  $   \sim \frac{r_{\rm Maj}^2}{ r_d^2} \frac{n_f}{n_i}  \approx \frac{r_{\rm Maj}^2 l^3}{ r_d^5} \frac{N_f}{N_i} $. This strongly depends on the dimple radius, a large radius being better in term of losses.

Taking into account two-body inelastic losses, as well as Majorana and three body atom losses we estimate the number of lost atoms to be: 
 $$N_{\rm loss}\sim N_f  \left[ t_{\rm coll} K_2 n_f + \frac{2}{3} t_{\rm coll} L_3 n_f^2 + \frac{r_{\rm Maj}^2}{ r_d^2} \frac{n_f}{n_i}   \right]$$
  where we choose $B\approx B' (\sigma_r)_d$ when evaluating the $K_2$ formula. Formula for the three body loss rate $\Gamma_3$, $t_{\rm coll}$ or $\Gamma_{\rm ev}$ are given in the next section.
 The number of loaded atoms 
  $N_f$ (calculated from the thermodynamical equation) might
never been achieved if the number of atoms lost $N_{\rm loss}$ is large, in  this case we are not able to calculate the number of loaded atoms but we can estimate it to be $N_f-N_{\rm loss}$.

\subsection{Numerical results}

In this section, we discuss the strategy to efficiently load a dimple trap, and give numerical results in the case of cesium. For our numerical simulations, we assume as before a $P=100\,$W Nd:YAG laser focused on a $w_0=100\,\mu$m waist. We also assume a $10\,$mT/cm gradient magnetic quadrupole trap containing $10^9$ atoms at a temperature of $150\,\mu$K. This corresponds to a final $\eta = -\frac{U_0}{k_B T_f}  \approx 8$ (different from $\eta_d= -\frac{U_0}{k_B T_i} =11.5  $) which is a convenient starting condition for evaporative cooling.

\begin{figure}
\resizebox{0.45\textwidth}{!}{
\rotatebox[origin=rB]{270}{
\includegraphics*[5mm,1mm][196mm,130mm]{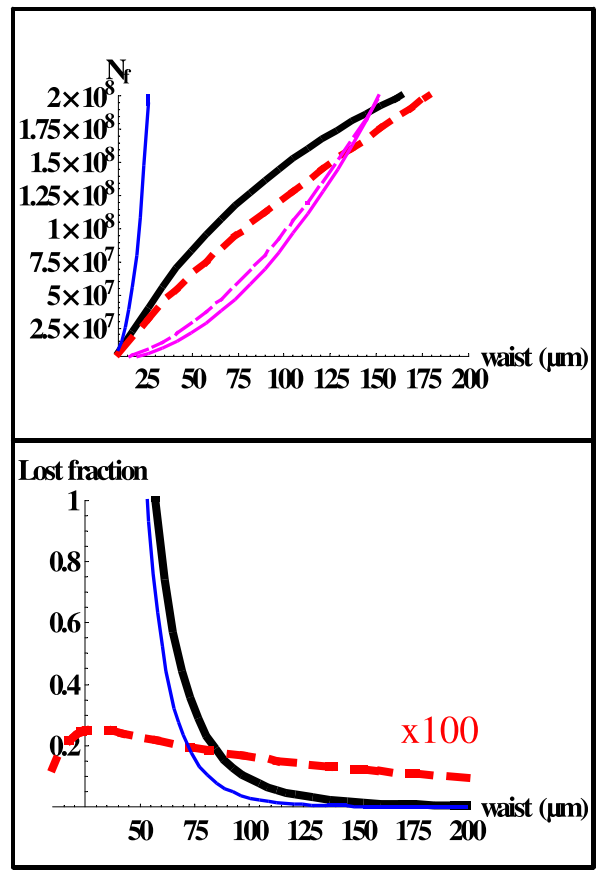}
}
}
\caption{Number of transfered atoms and losses during the loading of the optical trap  from the  reservoir containing $N_i=10^9$ atoms at temperature $T_i=150\,\mu$K. Top:
final number of transfered atoms $N_f$ resulting of our thermodynamical model where no 
losses are included.
 The red thick dashed lines represent  formula (\ref{eq_ana}), the thin blue solid lines are their approximations for small $\eta_V e^{\eta_d}$ and the thick black solid lines.   $N_d^0$, which is a naive  threshold atomic  for the hydrodynamical regime (see text), is also indicated with a thin magenta dashed line as well as its approximation $4 \frac{w_0^2}{\sigma}$.
Bottom: 
Lost fraction  $\frac{N_{\rm loss}}{N_f}$ of the loaded atoms due to three body relaxation (thin blue solid line), two-body inelastic collision (red dashed line) or Majorana transition (thick black solid line).
To be visible we have multiplied the negligible two-body inelastic losses by a factor 100.
}
\label{fig:dimple2}
\end{figure}

Based on thermodynamical considerations, as illustrated in figure \ref{fig:dimple}, one may believe that
a small volume dimple trap would insure a high initial phase space density and a small heating and can even be compatible with large atom numbers.
When three body losses,  Majorana losses and the hydrodynamical regime are taken into account the conclusion is quite different. This is especially true for cesium whose scattering length $a\approx -3000\,a_0$ is large at zero magnetic field.
It is essential to take into account Majorana losses, as well as two body and three body losses, which  only depend on the scattering length $a$.  For further evaporative cooling, one also needs to avoid the hydrodynamical regime, when the collisional probability $p_{\rm coll} \sim \frac{t_{\rm coll,d}}{t_{\rm osc,d}} $ is greater than one, i.e when energetic atoms collide before being evaporated. The strategy is then to choose a dimple radius $r_d$  in order to load $N_f \lesssim 2  N_i \frac{r_d^2}{l} $ atoms in a dimple  to avoid the hydrodynamical regime, while avoiding Majorana and inelastic losses, and then to start evaporative cooling.

Numerical results for loading  cesium atoms in a dimple are shown in figures 
 \ref{fig:dimple3}  and \ref{fig:dimple2}.  If from the pure thermodynamical point of view a small dimple trap looks to be the best choice,
 figure \ref{fig:dimple2} indicates that its loading is strongly affected by Majorana transitions  and three body losses and figure \ref{fig:dimple3} clearly indicates that the high density reached leads to the hydrodynamical regime and to such high three body recombination rate that it is problematic for subsequent evaporation.  
 Finally the (diabatic) loading of the dimple is fast.
We also conclude that in order to avoid three body losses and the hydrodynamical regime, and to have large number of atoms $ N_f \sim 4 w_0^2 / \sigma$ loaded, it is necessary to use a rather large waist. This has a cost in the final phase-space density, but, as we will see in the next section, evaporative cooling starting with such conditions is so fast that this remains the best strategy. In our case  we conclude that a $w_0=100\,\mu$m waist, $2\times 100\,$W laser (crossed laser of $100\,$W each),  is a good compromise. More than $10^8$ atoms can be transfered in hundreds of millisecond, leading to a phase-space density of $1/700$ before starting evaporation.

Our calculations show that the dimple loading due to collisions between atoms is a very efficient process which leads to very high phase-space densities compared to the ones in a MOT. In general, when $\eta_d\approx 4-12$, it is possible to find a waist value for which $20\,$\% of the atoms are loaded without significant heating,  and still avoiding the hydrodynamical regime as well as the Majorana and the two or three body losses. Such large atom numbers loaded in the dimple are an ideal starting point for evaporative cooling. After polarization of the atoms in their lowest state in energy, to avoid two-body inelastic losses, evaporative cooling is performed by lowering the optical trap depth.  This will be discussed in the following section.

\section{Evaporative Cooling}

In this section, we give a complete set of simple equations for evaporative cooling, and we discuss the optimal evaporative cooling strategies.  
Our theory generalizes the one developed in \cite{1996PhRvA..53..381L,2001PhRvA..64e1403O} (see also \cite{1997PhRvA..55.1281B,1997PhRvA..56.3308B}), by including the effects of gravity, three-body recombination events, the possible dynamical modification of the scattering length by a Feshbach resonance, as well as a time-dependent shape of the dipole trap. We derive simple differential equations for the atom number $N$, the temperature $T$ and the energy $E$ of the sample trapped in a time dependent trap, that can be  easily solved numerically.

We will assume an isotropic power law potential: $U(r) =U' r^{3/\delta}$ for $r<r_U$ 
($U(r_U)=\eta k_B T$ is the initial trap depth). 
The optical trap case  ($r_U = w/\sqrt{2}$) is idealized by the quadratic potential ($U(r) = \frac{1}{2} m \omega^2 r^2$ with $\delta=3/2$). For a non isotropic potential, similar equations can be written, for instance in the harmonic case  by taking the trapping angular frequency $\omega $ as the geometric average of the three different axis angular frequencies \cite{1996PhRvA..53..381L}.

We will focus on the high $\eta $ regime ($\eta>4$) where evaporative cooling is  more efficient (for evaporative cooling at low trap depth see \cite{2004PhRvA..70e3409D}). In this high $\eta$ regime, we assume that all the corrective factors (see \cite{1996PhRvA..53..381L}) such as 
$P_l(\eta) = 1 - \Gamma[l,\eta]/\Gamma[l]$  are equal to one, where $\Gamma$ is Euler's gamma function. This leads for instance to neglect the spilling terms in the evaporation equations.
We assume a Boltzmann gas behavior which is a good approximation except very close to the quantum degeneracy regime. For Bose gases, see \cite{1997PhRvA..56..560W,1999PhRvA..59.2243Y} and for degenerate Fermi gases, see \cite{2000PhRvA..61a3406G,2000PhRvA..61e3610H,2002PhRvA..65f3617G}.  We will assume that  quantum degeneracy is reached (for the real gas) when the phase space density $D$ reaches $1$ for the classical gas.

We assume  s-wave elastic scattering with an elastic collision rate $\Gamma_{\rm el} =  n_0 \sigma v_r$ where $v_r = 4 \sqrt{ \frac{ k_B T}{\pi m}} $ is the average atomic relative velocity, $\sigma = \frac{8 \pi a^2} {1+ (k a)^2}$ is the energy dependent scattering cross section with $ k = \frac{m v_r}{2 \hbar}  $. To take into account an average of the energy dependent scattering in the potential, the authors of 
 reference \cite{2004PhRvA..70a3404T} 
  modify the cross section formula. The equations (53)-(54) of reference \cite{2004PhRvA..70a3404T} are complex. Here, we modify them 
 by replacing $k^2$ by $k_{\rm ev}^2= k^2 \frac{\eta }{3} \frac{\pi}{4} $
 in the scattering cross section formula. We found that such an approximation is valid to an accuracy of 20 percent, and is more physically intuitive, as it explicitly introduces an average momentum $k_{\rm ev}$ in the new $\sigma = \frac{8 \pi a^2} {1+ (k_{\rm ev} a)^2}$ formula\footnote{Only when calculating collision in the 
 (infinitely deep)
 quadrupole  magnetic trap 
we still use $k_{ev}$ to replace $k$ but we choose in our calculation $\eta =4$,  and not $\eta=+\infty$,  to have $k_{\rm ev}^2 \approx k^2$ which restores the usual formula for the scattering cross section.}
For a three dimensional evaporation,  the evaporation rate is given  by $\Gamma_{\rm ev}^\eta = \Gamma_{\rm el} e^{-\eta} ( \eta - (\frac{5}{2} + \delta ) )/\sqrt{2}$ and the average energy taken by the evaporated atoms is $E_{\rm ev}^\eta \approx (\eta +1) N k_B T $. 
These formula are  15\% accurate compared to the ones containing the  $P_l(\eta)$ corrective factors \cite{1996PhRvA..53..381L}.

\subsection{Gravity effect}

\begin{figure}
\resizebox{0.45\textwidth}{!}{
\includegraphics*[11mm,79mm][135mm,203mm]{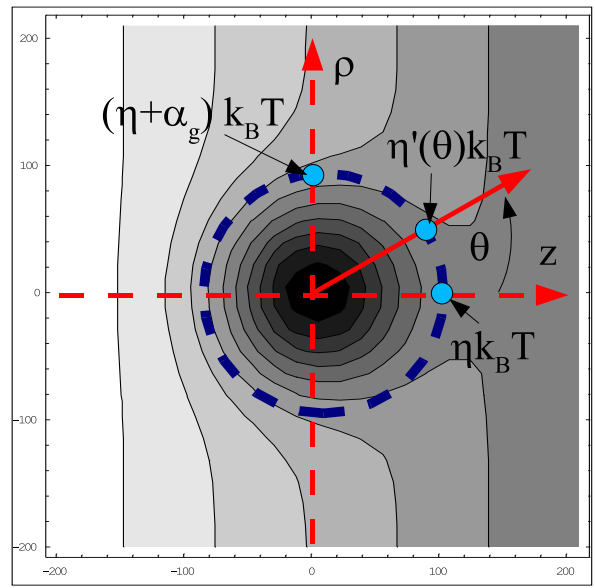}
}
\caption{Equipotential lines for an isotropic gaussian optical trap including gravity (corresponding to two crossed $5\,$W  Nd:YAG lasers  focused on  $w \approx 100\,\mu$m). Azimuthal $z$ and polar coordinates $\rho=\sqrt{x^2+y^2}$ in $\mu$m have been used (a part with the unphysical $\rho<0$ region is also drawn by symmetry for clarity). A ''trajectory'' with azimuthal angle $\theta$ is indicated crossing the evaporation (dashed) circle with $\eta'(\theta) k_B T$ minimal evaporation energy. The evaporation energies $\eta k_B T, \eta'(\theta) k_B T,( \eta+\alpha_g)k_B T$ are indicated respectively for $0,\theta,\pi/2$  angles.}
\label{fig:grav}
\end{figure}

Taking into account gravity, the full potential is $U_g(r,z)=U(r)-m g z$. The evaporation efficiency is substantially reduced  by the gravitational field when  $ \alpha_g= \frac{ m g r_U }{k_B T} > 1 $ \cite{Ketterle1996b,1998PhRvA..57.4747P,Davis1995a}. For $\delta<3$ the minimum of the potential $U_g^-$ is no more in $r=0$ and we have to redefine the trapping depth  by $\eta k_B T = U_g (r=r_U,z=r_U)-U_g^-$. To be used in the evaporative cooling equation $\alpha_g$ has to be written as an $\eta$ function. For instance,
in the harmonic trap case:
\begin{eqnarray}
\alpha_g &=& \frac{ m g  r_U}{k_B T} = \frac{ m g  }{k_B T} \left(\frac{g}{\omega^2} + \sqrt{\frac{2 \eta k_B T}{m \omega^2}} \right)  \label{eq:waist} \\
\omega  &= &\sqrt{\frac{4 \eta k_B T}{m w_0^2}} \left( \frac{1}{2} +  \frac{1}{2} \sqrt{1+\frac{\sqrt{2} m w_0 g}{\eta k_B T}} \right). \nonumber
\end{eqnarray}
For the crossed dipole trap case, $w_0 =  \sqrt{2} r_U $, $\omega =\sqrt{\frac{4 U_0}{m w_0^2}}$ and $ U_0\propto P/w_0^2 \propto \omega^2 w_0^2$ expressions are still valid (but $U_0=U(r_U)$ is no more the trap depth $\eta k_B T$ except in the zero gravity case) .

During a typical optical evaporation process the laser power is reduced and the effect of gravity increases.
Gravity tilts the potential, and modifies the evaporation rate, as is illustrated in figure \ref{fig:grav}. For the sake of simplicity, here, we only consider the case of isotropic traps, and we only describe ''central'' ($r=0$) collisions. After such a collision, a given atomic trajectory oscillates between $z<0$ and $z>0$ so atoms will escape toward the $z>0$ region where the potential is the lowest. We label the trajectory with the azimuthal angle $\theta$ (such as $z =r \cos \theta$), between $0$ and $\pi/2$. 
 Due to isotropy the density probability of the azimuthal angle between $\theta$ and $\theta + d \theta$ is $ \sin \theta d \theta$.
 The evaporate atom energy is mainly  between the threshold energy $ \eta'(\theta)  k_B T =
 U_g(r=r_U,z=r_U \cos \theta)-U_g^-$
   and $ (\eta'(\theta) +1) k_B T $ \cite{Ketterle1996b}. Therefore, and
for simplicity reason, we assume that the $\theta$ angle is determined exactly by the evaporation energy.
 Then we choose to average the escape rate to $\Gamma_{\rm ev} = \int_0^{\pi /2}\Gamma_{\rm ev}^{\eta'(\theta)} \sin \theta  d \theta$ and the average energy  to $E_{\rm ev} =  \int_0^{\pi /2}  \frac{\Gamma_{\rm ev}^{\eta'(\theta)}}{\Gamma_{\rm ev}}  E_{\rm ev}^{\eta'(\theta)} \sin \theta  d \theta$, which with $\eta'(\theta) =\eta + \alpha_g(1- \cos \theta)$ are:
 \begin{eqnarray*}
 \Gamma_{\rm ev} & = & \frac{ \Gamma_{\rm el} e^{-\eta - \alpha_g} }{\sqrt{2}} \left[ \left( \eta - (\frac{3}{2} + \delta) \right) \frac{e^{\alpha_g}-1}{\alpha_g} - 1 \right]  \\
 \frac{ E_{\rm ev} - E_{\rm ev}^\eta }{N k_B T}  & = &  \frac{  e^{-\eta-\alpha_g } }{ \Gamma_{\rm ev}} \left[ \left( \eta - \frac{1}{2} - \delta \right) \left( \frac{e^{\alpha_g}-1}{\alpha_g} - 1 \right) - \alpha_g \right] 
 \end{eqnarray*}
 We could easily check that the limit of no gravity $\alpha_g \rightarrow 0$ is correct. These formula are probably not accurate for strong gravity effects, but they will be used to test if the gravity is negligible in a given evaporative cooling scheme. 

 \subsection{Hydrodynamical effects}

  Evaporation is efficient only in a non hydrodynamical regime where the atoms do not collide before reaching the evaporation point \cite{Zy2004}. The probability $p_{\rm coll}$ for colliding before the evaporation point  can be estimated by dividing the typical size of the cloud $\ell$ by the mean free path $(\bar n \sigma)^{-1}$   where $\bar n = n_0 2^{-\delta}$ is the average atomic density. In the harmonic trap case $\ell \approx \sqrt{\pi} \sigma_r$  \cite{2000PhRvA..62f3614B,2001PhRvL..87q0404S},
leading to the approximate formula  $p_{\rm coll} \approx \frac{\Gamma_{\rm el}}{4 \omega}$. This is similar to the  crossover criterium between the hydrodynamical and the Knudsen regime given by \cite{1997PhRvL..78.1838G}. We then approximately take into account the collisional chain  by multiplying the atom and the energy losses by an empirical smoothing function  $f (p_{\rm coll}) = \left(1+ p_{\rm coll}^3 \right)^{-1}$ which smoothly tends to zero in the high hydrodynamical regime. We believe that our results are quite insensitive to the exact function which is chosen to describe hydrodynamical effects because we tried to remain in situation where these effects are negligible.

\subsection{Three body recombination}

Three body recombination (TBR)  leads to losses at a rate $\Gamma_3 = L_3 \overline{n^2} = L_3 n_0^2 3^{- \delta}$. The detailed expression is complex and depends on the molecular potential. Here, we will
 use the upper bound expression:  $L_3 \approx 225 \frac{\hbar}{m} \frac{a^4}{1+ 0.1 (k a)^4} $
 which is an approximate formula based on the experimental results of reference \cite{2003PhRvL..91l3201W} and on theoretical works \cite{2004NuPhA.737..119G,2004PhRvL..93l3201D,2005PhRvL..94u3201D,2005PhRvA..72d3607Z}. 
 This expression is valid if no resonance processes are present (such as Efimov states occurring for $a<0$ at very low temperature \cite{2005cond.mat.12394K}). It leads (we use $k_{\rm ev}$ instead of $k$ in $L_3$ formula) to a  simple result: 
 $L_3 \lesssim 3.9 \frac{\hbar}{m} \sigma^2 $ and therefore
 
 \begin{equation}
 \Gamma_3 \lesssim 0.15 \times 3^{ 3/2-\delta} \frac{\hbar \Gamma_{\rm el} }{k_B T} \Gamma_{\rm el}. \label{TBR}
 \label{TBR_rate}
 \end{equation}
Moreover, TBR leads to ''anti-evaporation'' (the loss of the coldest atoms). A spatial averaging of the potential energy $U(r)$ with the TBR rate $ \Gamma_3$ leads to an energy heating of $\frac{2}{3} \delta k_B T$ per event. TBR formes molecules in their highest vibrational state, with binding energy $k_B T_h = \frac{2 \hbar^2}{3 m a^2}$ (formula valid for large scattering length value). 
We will not take into account complex possible ro-vibrational relaxation. But, we have to take into account the fact that the
atomic (or molecular) products  of TBR are trapped in the sample, which is therefore heated by $T_h$ per event, if $k_B T_h < \eta k_B T$. \cite{2003PhRvL..91l3201W}. This heating does not occur if $k_B T_h>\eta k_B T$ because these particles are ejected from the trap, except if the hydrodynamical regime is reached. To take into account this possibility, we consider that TBR leads to a heating term of $ k_B T_h$ multiplied by a smoothing function $f_{\rm TBR}  =  1-f(p_{\rm coll}) \left(1- f \left( \frac{k_B T_h}{\eta k_B T} \right) \right)$ which is $1$ in all cases except when $k_B T_h>\eta k_B T$ and in a non hydrodynamical regime.
 
 \subsection{Evaporative cooling equations}
 
 We consider a generic loss term $\Gamma_{\rm loss} = \Gamma_{\rm in} + \Gamma_{\rm bg} $ where  $\Gamma_{\rm bg}$ is the background collisional rate and $\Gamma_{\rm in} =K_2 \bar n $ is the
inelastic loss rate.  $\Gamma_{\rm in}$ can be zero in an optical trap where the atoms can be trapped in their true ground state. The energy $E$ of the atoms
is  modified by the change of the trap shape through the change of potential energy $E_{\rm pot} = \delta N k_B T= E \frac{ \delta}{3/2+\delta}$  occurring at a rate $\Gamma_{\rm pot} = \frac{\dot U'}{U'} $ \cite{1997PhRvA..55.1281B,1997PhRvA..56.3308B,2001PhRvA..64e1403O}.
 In an optical trap  $E$  can change by $N k_B$ times the recoil temperature $T_{\rm recoil}$ due to photon absorption at a rate $\Gamma_{\rm laser}\propto \eta k_B T$.  
Finally, the evaporative cooling equations to be solved are 
\begin{eqnarray}
\dot N& = & - \left[ \Gamma_{\rm ev} f(p_{\rm coll})   + \Gamma_{\rm loss} + \Gamma_3 \right] N  \label{eq:evaporat} \\
 \dot  E& = &  - \left[ \Gamma_{\rm loss} +  \Gamma_3 \right] E - E_{\rm ev} \Gamma_{\rm ev} f(p_{\rm coll})   + E_{\rm pot} \Gamma_{\rm pot} + \nonumber \\
\lefteqn{  N  \left[ \Gamma_{\rm laser} k_B T_{\rm recoil} + \Gamma_3 ( \frac{2}{3}\delta k_B T + k_B T_h f_{\rm TBR} )\right]    } \nonumber 
\end{eqnarray}
We stress that these equations take into account all relevant physical phenomena for evaporative cooling: temperature dependent cross-section, one-body (such as background collisions) and two-body losses, TBR, hydrodynamical regime, gravity, arbitrary modification of the trap depth and shape, time-dependent scattering length. All terms in this set of equation can be written as a function of $N$, $T$,
 $a$ and $U$ (i.e. $\eta$ and $\omega$ in the harmonic case).
Considering that
 $\frac{\dot E}{E} = \frac{\dot N}{N} + \frac{\dot T}{T} $ these equations lead to differential equations for $N$ and $T$ which only depend on
 $a$ and $U$. Simple analytical solutions exist in simple cases, when there is no three body process, no gravity, $\sigma$ is not temperature dependent... In our case, these equations need to be solved numerically, and we use a simple Mathematica program. These programs, as the one describing the dimple loading, are available upon request to the corresponding author.

\subsection{Evaporative cooling strategy}

The  optimization of evaporative cooling is usually achieved  by maximizing   at each time 
 $\gamma= \frac{\dot D/D}{\dot N/N} $ ($\gamma>3$ for good experimental conditions \cite{2004ApPhB..79.1013K,2005cond.mat..8423G}). But, if there is no inelastic, TBR or background gas processes, $\gamma$ is maximized for an infinite $\eta$ value which leads to infinite evaporation time \cite{Ketterle1996,1997PhRvA..55.3797S}.
 A better parameter, at least for the speed of the evaporation process,
is  $\frac{\dot D}{D}=\frac{\dot N}{N}-(\frac{3}{2}+\delta)\frac{\dot T}{T}+\delta \Gamma_{\rm pot}$.
More fancy parameters such as $ \frac{(\dot D/D)^2}{\dot N/N}$ can also be used to optimize at each time the evaporative cooling process.
$\frac{\dot D}{D}$
 becomes, in the zero gravity case,
\begin{eqnarray}
\frac{\dot D}{D} &=& \frac{\Gamma_{\rm el}}{\sqrt{2}} e^{-\eta} ( \eta - \frac{5}{2} - \delta )(\eta-  \frac{3}{2} - \delta)  f(p_{\rm coll}) -\Gamma_{\rm loss} - \nonumber
\\
& &
\Gamma_{\rm laser} \frac{T_{\rm recoil}}{T} - \Gamma_3 \frac{\frac{5}{3}\delta T + T_h f_{\rm TBR} }{ T}.  \label{evapD}
\end{eqnarray}
$ \Gamma_{\rm pot}$ does not appear in this equation because  changes in the potential shape do not affect the phase space density. This formula is valid (as well as equations (\ref{eq:evaporat})) when the adiabatic conditions $\frac{\dot  U_0}{ U_0} \ll \Gamma_{\rm el}$ and $ \frac{\dot  U'}{ U'} \ll \frac{\omega}{2 \pi}   $ (in the harmonic case) \cite{1997PhRvA..55.1281B,2004PhRvA..70a3615M}
 are verified. We have numerically verified that all the results presented in this article follow these adiabatic conditions within $90\%$.
Equation (\ref{evapD}) indicates that
$\dot D/D$ is maximized for $\eta \approx 4.1 + \delta $, high $\Gamma_{\rm el} $ just before  the hydrodynamical regime (optimal value $\Gamma_{\rm el} \approx 3 \omega $ in the harmonic case) and negligible loss rate. We have as typical value
$\frac{\frac{5}{3}\delta T + T_h f_{\rm TBR} }{ T} \approx \frac{5}{3} + \eta  \approx 10$. Therefore, 
 the TBR heating rate  is $\sim 10 \Gamma_3$.  We then see, from equation (\ref{evapD}), that in order to have an efficient evaporation
the TBR heating rate $\sim 10 \Gamma_3$, the loss rate $\Gamma_{\rm loss}$, and the laser absorption rate $\Gamma_{\rm laser} T_{\rm recoil}/T$ must not be higher than $\Gamma_{\rm ev}$. 
 Using
equation (\ref{TBR_rate}) and  (in the harmonic case) $ \Gamma_{\rm ev} ^{\eta=6}  = 0.0035 \Gamma_{\rm el}$  we found that the TBR term is negligible if
$\Gamma_{\rm el} (s^{-1})  \ll 300 T(\mu {\rm K})$.
 An efficient evaporation should then verify for $\eta \approx 6$
\begin{equation}
\Gamma_{\rm el} (s^{-1}) \approx \omega  \ll 300 T(\mu {\rm K}) \label{TBR:limit}
\end{equation}
For $\eta =10$ this becomes $\Gamma_{\rm el} (s^{-1}) \ll 15 T(\mu {\rm K})$ which is much  harder to achieve.
This equation already indicates that, in contrary to the common intuition, the choice of a high initial temperature is not a bad choice because it  allows to have a tight trap with fast evaporation due to high collisional rate without reaching the  three-body collisional regime.

 A very crude estimate of the minimum time $t_{\rm BEC}$ needed to reach the degeneracy starting from an initial phase space density $D_i$ can be estimated from equation (\ref{evapD}) with no losses and
 $ \Gamma_{\rm el} \approx  \omega $ kept constant during the evaporation process leading to $t_{\rm BEC} \approx - \frac{\ln (D_i)}{0.01 \omega}$  
 for $\eta \approx 6$ or $t_{\rm BEC} \approx - \frac{\ln(D_i)}{0.001 \omega}$ for $\eta \approx 10$.
 $\eta \sim 10$ can be used to improve the final number and $\eta =6$ is a faster strategy to reach BEC.

 \subsection{Numerical results}

Before we detail the conclusions of our study, in this paragraph, we compare our numerical simulation to the experiment described in reference \cite{2001PhRvL..87a0404B} and modeled  in reference \cite{2001PhRvA..64e1403O}. We choose the parameters  of
the table 1 of reference \cite{2001PhRvA..64e1403O}: $m$ is here the
 rubidium mass, $a=100\,a_0$, $N_0=6.7 \times 10^5$ initial atom number, $T_0=38\mu K$ initial temperature, $\omega_0 = 2\pi \times 1500\,$Hz initial angular frequency, $\Gamma_{\rm loss}^{-1} =6\,$s and $\eta=10$.
 Formula (\ref{eq:waist}) lead to an effective waist of $w_0= 40\,\mu$m  slightly different than the  experimental waist of $\lesssim 50\,\mu$m. 
  
\begin{figure}
\resizebox{0.45\textwidth}{!}{
\rotatebox[origin=rB]{270}{
\includegraphics*[45mm,50mm][139mm,178mm]{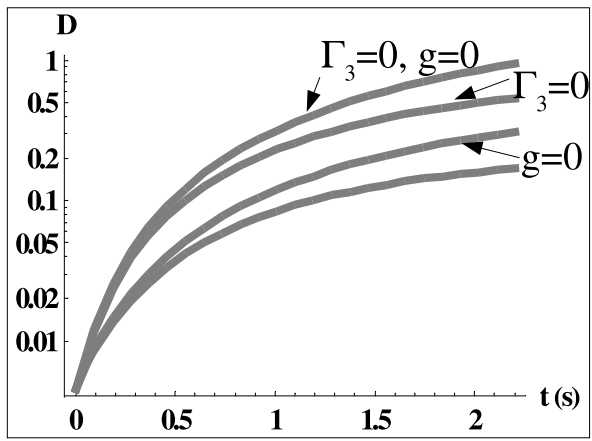}
}
} 
\caption{Evolution of the phase space density during the evaporation at $\eta=10$ and constant laser waist (see text). 
The four curves represents respectively, from the bottom one to the top one, full numerical resolution of the equations (\ref{eq:evaporat}), resolution  with no gravity ($g=0.01\,$m$^2$/s), without taking into account the $\Gamma_3$ term, and with no gravity neither three body loss terms.}
\label{fig:evap_eta}
\end{figure}

 Our results are shown in figure \ref{fig:evap_eta}. The hydrodynamical regime is never reached. When both gravity and TBR are neglected, we recover the theoretical results of \cite{2001PhRvA..64e1403O}. However, when we take those effects into account, we predict that the choice of parameters of reference \cite{2001PhRvA..64e1403O} would not lead to BEC. We found that it is nevertheless possible to reach BEC, in agreement with the experimental observation \cite{2001PhRvL..87a0404B}, starting the evaporation with the experimentally measured parameters $N_0=2 \times 10^6$ and $T_0=75\, \mu K$.  Our numerical treatment is therefore able to reproduce the experimental observations, which theories without gravity or TBR cannot do as accurately.

 \subsection{Evaporation of the dimple optical trap}
 
We now turn to our numerical results and discuss the best strategy to reach BEC in the case of cesium. We start evaporation in conditions close to the one deduced from our dimple loading theory
with a $100\,$W Nd:YAG laser focused on $100\,\mu$m:
 $N_0=10^8$, $T_0=200\,\mu$K, $\eta=9$. Such a dimple trap has a  heating rate of $\Gamma_{\rm laser} \times T_{\rm recoil} \approx 11\,$s$^{-1} \times 0.2 \,\mu$K at full power. This heating rate will be found to be negligible.
All parameters $\eta,\omega,a,g$ can be experimentally adjusted, and our theory takes this into account. 
In particular, gravity $g$ can be effectively suppressed by adding a vertical levitating magnetic field gradient on the order of tens of mT/cm \cite{2005PhRvA..71a1602K}.
 This levitating magnetic field gradient 
also produces a parabolic anti-trapping potential, with frequencies on the order of a few Hz,
which is usually a small effect that we will neglect \cite{2004ApPhB..79.1013K}.

\begin{figure}
\resizebox{0.49\textwidth}{!}{
\rotatebox[origin=rB]{270}{
\includegraphics*[7mm,7mm][184mm,184mm]{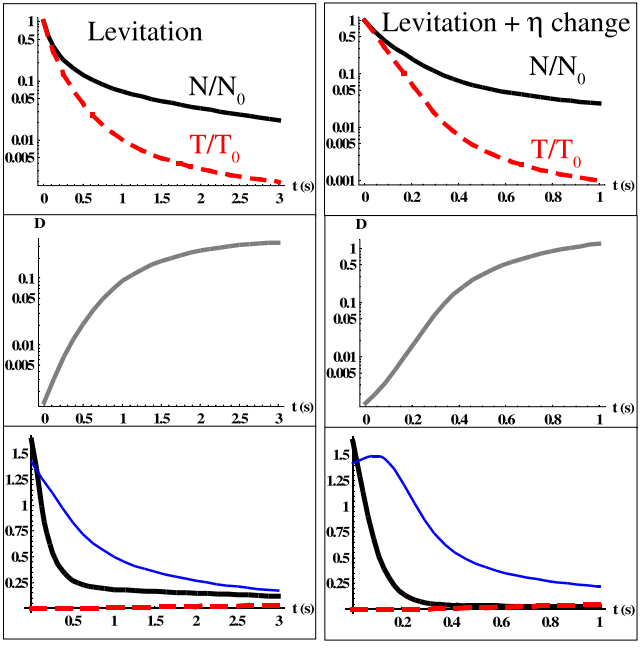}
}
} 
\caption{Efficiency of the evaporative cooling from numerical resolution of the equations (\ref{eq:evaporat}) for constant waist, without gravity ($g=0.01\,m^2/s$), $a=100\,a_0$, $N_0=10^8$, $T_0=200\,\mu$K. Left: $\eta=9$, Right:
$\eta$ decreases  in a $0.2\,$s exponential decay from $\eta=9$ towards an asymptotic value of $\eta=6$.
Top: evolution of the atom
number N (solid line) and the temperature T 
(dashed line). Middle: evolution of the phase space density $D$. Bottom:  hydrodynamical  $p_{\rm coll}$  (thin blue
solid line), three body $\Gamma_3/\Gamma_{\rm ev}$
(thick black solid line) and 
gravity $\alpha_g$ (dashed red thick line) dimensionless parameters.}
\label{fig:evap_eta_Levi}
\end{figure}

\begin{figure}
\resizebox{0.45\textwidth}{!}{
\rotatebox[origin=rB]{270}{
\includegraphics*[7mm,19mm][198mm,194mm]{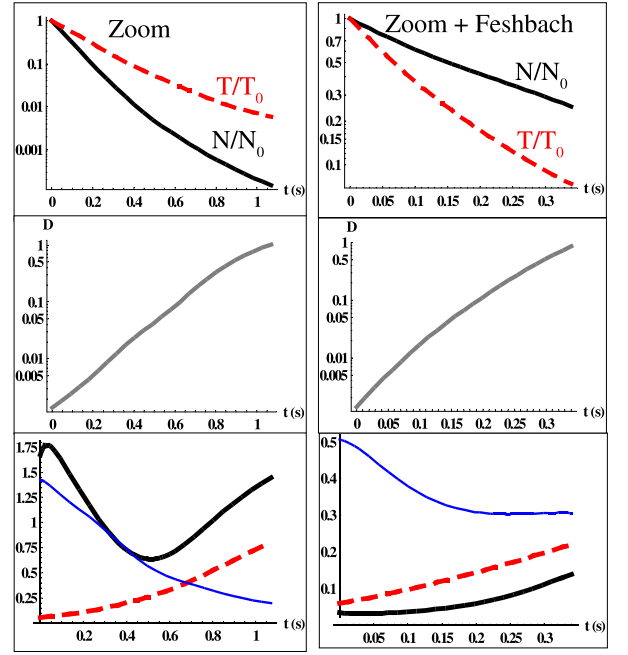}
}
} 
\caption{Efficiency of the evaporative cooling from numerical resolution of the equations (\ref{eq:evaporat}) for a waist zoom, chosen here to keep the trapping frequency $\omega$ constant, $N_0=10^8$, $T_0=200\,\mu$K, $\eta=9$. Left:  $a=100\,a_0$ and right: $a=30\,a_0$ decreases in a $0.2\,$s exponential decay towards an asymptotic value of $10\,a_0$. Top: evolution of the atom
number N (solid line) and the temperature T 
(dashed line). Middle: evolution of the phase space density $D$. Bottom:  hydrodynamical  $p_{\rm coll}$  (thin blue
solid line), three body $\Gamma_3/\Gamma_{\rm ev}$
(thick black solid line) and 
gravity $\alpha_g$ (dashed red thick line) dimensionless parameters.}
\label{fig:evap_eta_var}
\end{figure}

As illustrated in figure \ref{fig:evap_eta}, a strategy with constant waist ends up with gravity issues, due to quasi one dimensional evaporation.
We illustrate different evaporative cooling strategies in figure \ref{fig:evap_eta_Levi} using a  levitating magnetic field  but a constant waist to avoid the gravity problem and in figure \ref{fig:evap_eta_var} using a time changing waist. We assumed losses due to collisions with hot atoms from the background gas, leading to a maximum lifetime of $3s = \frac{1}{\Gamma_{\rm loss}}$. We choose a scattering length value of $100\,a_0$ because it is found to be close to the optimal value and, despite the mass difference, it can mimic the rubidium case. 
 Even in the  levitating magnetic field  case, BEC is not reached if $\eta=9$ is kept constant. By rapidly reducing $\eta$ to a value $\eta = 6$, three body recombination and hydrodynamical regime are avoided, and BEC is reached in one second. This picture points out the effects of TBR and hydrodynamical regime, which our theory can predict and help avoiding. 
 
In the left part of figure \ref{fig:evap_eta_var}, we show how to reach degeneracy in less than one second by dynamically changing the waist of the dimple. The strategy chosen here is to keep the 
trapping frequency $\omega$ constant (see equation \ref{eq:waist}),
In this case, the condensated atom number is very small, because of hydrodynamical regime or three body losses. However, if $a$ is dynamically modified, as illustrated by the right curves in figure \ref{fig:evap_eta_var}, BEC can be reached in much less than one second, with large atom numbers. A Feshbach resonance is used to have a high collisional rate $\Gamma_{\rm el}$. However, this rate is kept lower than $3 \omega$ to avoid the hydrodynamical regime and lower than $
 40 \times  T(\mu{\rm K})$  to avoid TBR (the factor $40$ depends on the  $\eta$ value chosen, here $\eta\approx 9$). We reach
   degeneracy in a $0.35\,$s evaporation ramp with $N=2 \times 10^7$ final atom number at a temperature of $14\,\mu$K. During this time the waist is divided by a factor $3$.

To find 'the' optimized strategy to reach BEC by dynamically changing all $P,w,a$ or gravity parameters is beyond the scope of this paper. However, our studies show that keeping $\Gamma_{\rm el} \approx  \omega \approx
  100 T(\mu{\rm K})$ for $\eta\approx 8$  produces even bigger degenerate samples in less than $100\,$ms, by maximizing the collision rate, while avoiding the hydrodynamical regime.

\section{Conclusion}

In this paper, we have studied the sudden superimposition of a dimple trap on a large volume trap. We propose three schemes to experimentally realize this goal: a dark-SPOT, a (aberrated) dipole trap, or a magnetic trap as the atom reservoir. 
The dynamics of the loading leads to an exponential grow of the atoms number in time and the diabatic loading is three times faster than the adiabatic one.
The optimal dimple trap is the tightest possible trap  before reaching the hydrodynamical or the three body recombination regime allowing an almost temperature invariant  diabatic loading, i.e. with a maximum of 20\% of the atoms loaded. 
We have shown that the analytical equations (\ref{eq_ana})  can be used to find the right range for the 
dimple volume and radius. 
The focus of our study is a magnetic reservoir and a laser crossed dipole trap for the dimple. 
The magnetic setup is very simple because only two independent coils can provide magnetic field for the MOT, the quadrupole reservoir, the levitation  and even provide the field for a Feshbach resonance.
A linear quadrupole trap can be chosen to avoid two body decay but the size of the dimple trap has to be large enough to avoid the effect of Majorana losses which have been studied. Spin flip process can also be avoided  if the crossed optical trap is slightly off axes, where the magnetic field is not zero, or by using a quadratic magnetic trap. 
In this case of a dimple created by crossed lasers the 
laser power and waist $w_0$ have to be chosen to load $\sim 4 \frac{w_0^2}{ \sigma }$  (where $\sigma$ is the scattering cross-section) atoms from a magnetic trap in a quasi hydrodynamical regime. 

We then studied in detail teh evaporation inside the dimple trap.
For the first time, to our knowledge, we have derived simple evaporation equations including two-body, three-body inelastic collisions, hydrodynamical regime, effect of gravity, possible dynamical modification of both the temperature-dependent scattering cross-section and of the trap parameters.
  These equations  (\ref{eq:evaporat})
 can be extended to anisotropic potentials, low dimensional gases or to gases with  anisotropic electrostatic interaction \cite{2004PhRvA..69a2706M}. 
We discussed the best evaporative cooling strategy for rapidly reaching quantum degeneracy with large atom numbers. We suggest to use a tight trap, and an  elastic collisional rate $\Gamma_{\rm el}$ as high as possible but 
not higher than $
  300 T(\mu{\rm K})$ (here $\eta\approx 6$, but the factor $300$ depends on the  $\eta$ value chosen) to avoid the three body collisional losses  and not higher than $ \omega $ (in the harmonic trap case) to avoid the hydrodynamical regime. 
   The $\eta$ parameter could be chosen between $ 6$, for fast evaporation, and more than $10$ for large final number of atoms. We described how a levitating magnetic field gradient could be used  as well as dynamical modifications of the scattering length and the trap shape (laser waist for instance) to lead to a very fast evaporation and large final atom numbers. 
  Our main conclusion, which also holds for
Fermi or Bose mixtures \cite{2004PhRvA..70f3614B,2005PhRvL..95q0408S,2005PhRvA..72c3408A}, is that, 
using a tight trap with high frequency $\omega$ and high collisional rate $\Gamma_{\rm el}$
   but low scattering length and  high temperature,
  large ($> 10^7$ atoms) degenerate samples could then be achieved in much less than one second starting from a standard MOT.
  In our group an experimental activity is under way to test these schemes.

Acknowledgments: This work is in the frame of the 
Institut francilien de recherche sur les atomes froids (IFRAF) and of the
European Research and Training
Networks COLMOL (contract HPRN-CT-2002-00290) and QUACS (HPRN-CT-2002-00309).

A. F. acknowledges R. Grimm and H-C N\"{a}gerl and P. Lett for helpful discussions.

 B. L. T. acknowledges financial support from Conseil R\'{e}gional d'Ile-de-France, Minist\`{e}re de l'Education, de l'Enseignement Sup\'{e}rieur et de la Recherche, and European Union (FEDER - Objectif 2).

\end{document}